# Understanding and minimizing ac losses in CORC cables of YBCO superconducting tapes

**Linh N. Nguyen[1], Nathaniel Shields[2], Stephen Ashworth[2], and Doan N. Nguyen[1]**

[1] Los Alamos National Laboratory, Los Alamos, NM 87544 USA.
[2] VEIR, Boston, MA 02210, USA.

E-mail: xxxxxx

## Abstract

AC losses in conductor-on-rounded-core (CORC) cables of YBCO high-temperature superconducting (HTS) tapes are a significant challenge in HTS power applications. This study employs two finite element analysis (FEA) models to investigate the contributions from different AC loss components and provide approaches for reducing AC losses in cables. An FEA model based on T-A formula treats the cross-section of thin superconducting layers as 1D lines and, therefore, only can predict the AC loss generated by the perpendicular magnetic field. In contrast, the model based on H-formulation can be performed on the actual 2D rectangular cross-section HTS tapes to provide the total AC losses generated by magnetic fluxes penetrating from both the edges and surfaces of HTS tapes, although this model requires more computing time and memory. Both 1D and 2D simulation approaches were employed to offer a comprehensive understanding of the effects of cable design and operational parameters on the AC loss components in a 2-layer CORC cable. The research results given in this paper are therefore not only valuable to suggest strategies for reducing AC loss in multi-layer cables but also for developing more accurate and effective methods to calculate AC loss in CORC HTS cables.

Keywords: CORC cable, Superconducting cables, AC losses, Surface loss, Edge loss, Finite element modeling,

## 1. Introduction

The adoption of conductor-on-rounded-core (CORC) cables of YBCO high-temperature superconducting (HTS) tapes has great potential to enhance the efficiency and capacity of power transmission, yielding more reliable and sustainable electrical grid infrastructure [1-5]. In CORC cables, superconducting tapes are helically wound around a round core to form a compact, mechanically-robust cable structure that can offer high current capacity and excellent flexibility. The helical winding configuration also enables improved current sharing among the tapes, reducing the risk of local hot spots and contributing to a more uniform current distribution. However, AC losses in those HTS cables are still an important technical challenge because they reduce cable efficiency and generate thermal instabilities that may damage the HTS cable. Understanding and reducing AC loss in HTS cables are therefore essential and have been of great research interest, with remarkable efforts to evaluate and predict AC loss in CORC cables by both experimental [6-13] and numerical approaches [14-34].

AC losses in YBCO tapes are generated by the oscillating movements of flux lines during AC cycles. Due to the very high aspect-ratio geometry of YBCO tapes, the magnetic fields applied perpendicular to the wide surface of the tapes (perpendicular magnetic field) generate much higher AC loss than the magnetic fields applied parallel to the wide surface of the tapes (parallel fields). YBCO tapes produced with thin substrates (~30 μm) enable the construction of CORC cables with well-conformed tapes, as seen in Fig. 1, which presents the cross-section of a CORC cable [1, 35]. CORC reduces the perpendicular magnetic fields applied to the tapes; the field is primarily circumferential. Perpendicular fields only arise in regions near the gaps between tapes (or near the edges of the tapes). We distinguish these losses as the surface loss, generated by flux penetration of the parallel field, and the edge loss, generated by flux penetration of the perpendicular fields. Similar nomenclature was introduced in [16] for a CORC cable and [17] for non-inductive coils.

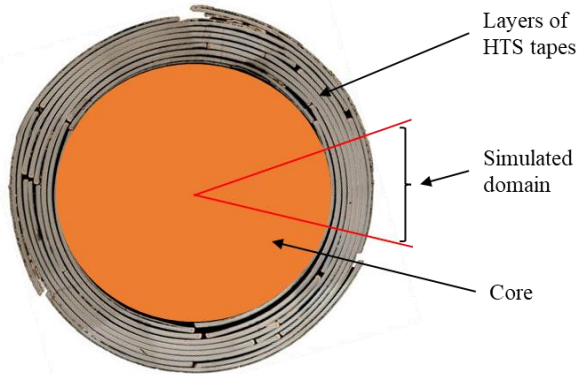

Fig. 1. Cross-section of a typical YBCO CORC superconducting cable, consisting of a metal core and HTS tape layers wound around the core [1, 35].

To account for the spiral geometry of the CORC cables when calculating AC losses in CORC cables, 3D FEM COMSOL Multiphysics [36] simulations have been used [18-21]. Those 3D models, however, must artificially increase the modeled thickness of HTS by tens of times to reduce the mesh and computational costs to manageable levels. Such treatment overestimates surface loss and results in a considerable error for scenarios in which surface loss dominates. AC loss can also be predicted by 2D FEM simulations performed on cable cross-sections. Those simulations can be either 1D models, which treat the HTS layers as lines [22-26] or 2D models based on H-formulation [27-32]. The 1D models can only calculate edge losses. The 2D models can compute total AC losses: edge and surface loss.

In this paper, both 1D and 2D simulation approaches are employed to relate cable design and operational parameters to the surface and edge losses of a 2-layer CORC cable by considering the (1) relative contributions of edge and surface losses to the overall AC losses; (2) impact of the tape alignment on AC losses in each HTS layer; (3) influence of the thickness of HTS layers on AC losses; (4) effect of size and number of inter-tape gaps on AC losses; and (5) contribution frequency on the AC losses. Our data are valuable for minimizing AC loss in multi-layer cables and improving the calculation of AC loss in CORC HTS cables.

## 2. Cable Design and finite element simulation models

We focus on two-layer cables. Most of our studies are done on Cable 1 in TABLE I, which consists of 8 tapes per layer. The cable $I_c$ and its former diameter are held constant throughout the studies to generate the same circumferential magnetic field on HTS tapes for all the studies. The width of the tapes is varied about 12 mm as required to adjust the inter-tape gaps. To understand the effect of the number of gaps on cable AC losses, Cable 2 (TABLE I) was used, similar to Cable 1, but with double the number of tapes, each with narrower width as detailed in TABLE I.

TABLE I. Properties of YBCO CORC superconducting cable

| Design | Cable 1 | Cable 2 |
|---|---|---|
| Number of layers | 2 | 2 |
| Number of tapes per layer | 8 | 16 |
| Former Diameter (mm) | 33.104 | 33.104 |
| $I_c$ of each HTS tape at 77 K (A) | 450 | 225 |
| Inter-tape gap, gap (mm) | 0.5, 1, 1.5, 2 | 0.5 |
| HTS tape width, $w_{HTS}$ (mm) | 12.5, 12, 11.5, 11 | 5.75 |
| Substrate thickness (μm) | 50 | 50 |
| Copper thickness (μm) | 30 | 30 |
| HTS thickness. $t_{HTS}$ (μm) | 1.5 - 10 | 1.5 |

A typical YBCO tape includes three electrically conductive constituents: the substrate, the superconducting layer, and the copper/silver stabilizer. All three are considered in our simulations. Fig. 2(a) illustrates the cross-section of a two-layer CORC cable. To reduce the computational workload, simulations only need to be performed in a subsector of the cross section by imposing periodic boundary conditions (Fig. 2a). We only consider AC losses generated in the YBCO tapes and ignore the losses generated in the core. Such an approximation is valid if the cable is designed to generate zero axial magnetic fields or the core is made with non-conductive material. The angle β is defined as the angle between adjacent gaps of the inner and outer HTS layers (see Fig. 2). With 8 tapes on each layer, β is bounded between 0° to 22.5°. Angle β = 0° represents the fully-aligned configuration and β = 22.5° represents the fully-misaligned configuration. Fig. 2(b) schematizes the arrangements of our simulations.

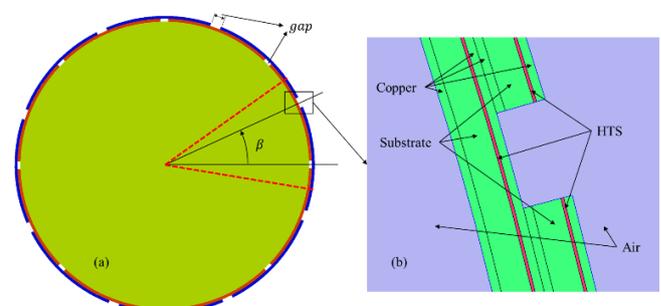

Fig. 2. (a) Cable cross section, illustrating the arrangement of HTS tapes. (b) Enlarged view of a gap between two tapes.

## 2.1 1D model (T-A formulation)

A cross-section of the HTS layer depicted in Fig. 2(b) is simplified to 1D (a curved line), as shown in Fig. 3. The T-A model requires the determination of two primary state variables: the current vector potential (**T**) in the superconducting layer and the magnetic vector potential (**A**) in the entire space [23-26].

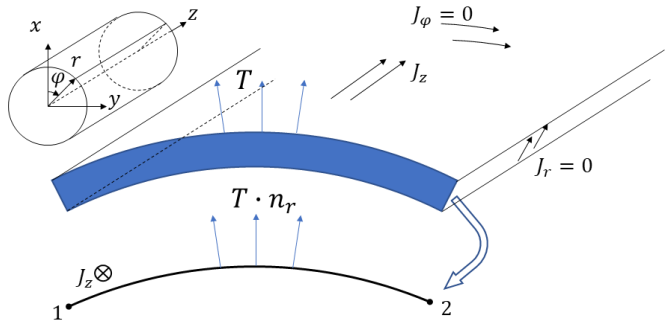

Fig. 3. In the T-A formulation, the thickness of HTS tape is ignored and the HTS layers were treated as 1D.

Within the superconducting layer, the current vector potential (**T**) is derived from the curl of the current density **J** and magnetic flux density **B**, as indicated by Eq. (1) and Eq. (2), respectively. To confine the computational HTS domain to 1D, the current is permitted to flow solely along the z axis; both $J_r$ and $J_\varphi$ are set to zero. The current vector potential is orthogonal to the HTS wide face at each point, allowing **T** to be computed as $T \cdot \boldsymbol{n_r}$, where $\boldsymbol{n_r}$ is the normal vector of the HTS curve, as formulated in Eq. (3).

$$\boldsymbol{J} = \nabla \times \boldsymbol{T}, \tag{1}$$

$$\nabla \times (\rho \nabla \times \boldsymbol{T}) = -\frac{\partial \boldsymbol{B}}{\partial t}, \tag{2}$$

$$\boldsymbol{n_r} = \left[ \frac{x}{\sqrt{x^2+y^2}}, \frac{y}{\sqrt{x^2+y^2}} \right]^T, \tag{3}$$

The superconducting characteristics of YBCO layers are characterized by the conventional power-law equation.

$$E_z = \rho J_z = \frac{E_c}{J_c(\boldsymbol{B})} \left( \frac{|J_z|}{J_c(\boldsymbol{B})} \right)^{n-1} J_z, \tag{4}$$

where $E_c = 10^{-4}$ V/m, $n$ is a constant and assumed to be 21. The critical current density $J_c(\boldsymbol{B})$ depends on both amplitude and direction of the magnetic field, as expressed in Eq. (5) below by [31,37].

$$J_c(\boldsymbol{B}) = \frac{J_{c0}}{\left(1+\sqrt{\left(B_\parallel k\right)^2 + B_\perp^2}/B_c\right)^b}, \tag{5}$$

where $B_\parallel$ and $B_\perp$ represent the parallel and perpendicular components of the magnetic flux density, respectively. $J_{c0}$ is the initial critical current density of HTS tapes. The parameters $k$, $B_c$ and $b$ take the following values 0.25, 0.1 T, and 0.7, respectively [31].

In the entire space, the magnetic field is determined using A formulation, as described in Eq. (6). The magnetic field distribution can be obtained from Eq. (7).

$$\nabla \times \left(\frac{1}{\mu} \nabla \times \boldsymbol{A}\right) = \boldsymbol{J}, \tag{6}$$

$$\boldsymbol{B} = \nabla \times \boldsymbol{A}, \tag{7}$$

Here **A** denotes the magnetic vector potential, **J** represents the sheet current density distribution in the HTS layer, and $\mu$ signifies the magnetic permeability.

By solving Eqs. (1-7), the electromagnetic properties of the HTS cable can be computed.

The integration of the current density across the cross-section of each HTS tape provides the total transport current flowing in that tape as describe the by Eq. (8) and reduced to 1D form as Eq. (9):

$$I_{HTS} = \iint_\Omega \boldsymbol{J} \, d\Omega = \iint_\Omega \nabla \times \boldsymbol{T} \, d\Omega = \oint_{\partial\Omega} T d\Omega, \tag{8}$$

$$I_{HTS} = (T_1 - T_2) t_{HTS}, \tag{9}$$

where $I_{HTS}$, $\Omega$, and $\partial\Omega$ are the transport current flowing in the HTS layers, their cross-section, and their boundaries, respectively. $T_i$ (i=1,2), are the current potentials at boundary points 1 and 2 as seen Fig. 3. Thus, Eq. (9) was used as the constraint in T-A models.

## 2.2 2D model (H formulation)

In the 2D model, H-formulation [27, 32] is employed. The governing equations of this model include Ampere's Law (Eq. (10)), Faraday's Law (Eq. (11)), the Constitutive Law (Eq. (12)), and the E-J power Law (Eq. (4)). Notice that the $J_c(\boldsymbol{B})$ in Eq. (5) is applicable to both our 1D and 2D models.

$$\boldsymbol{J} = \nabla \times \boldsymbol{H}, \tag{10}$$

$$\nabla \times \boldsymbol{E} = -\frac{\partial \boldsymbol{B}}{\partial t}, \tag{11}$$

$$\boldsymbol{B} = \mu \boldsymbol{H}, \tag{12}$$

From Eqs. (4) and (10)-(12), we find the basic relation

$$\nabla \times \rho \nabla \times \boldsymbol{H} = -\mu \frac{\partial \boldsymbol{H}}{\partial t}, \tag{13}$$

Similarly, as in our 1D models, the integration of current density over the cross-section of each HTS tape domain should provide the transport current flowing in that tape as seen in Eq. (14). This equation is imposed in weak constraints in our 2D models.

$$I_i = \iint_{\Omega_i} J_z \, d\Omega_i, \tag{14}$$

where $I_i$ and $\Omega_i$ (with $i$ = inner or outer layer) are the transport currents and domains of the inner or outer HTS layers, respectively.

## 2.3 AC loss calculation

Simulations are performed for 1.5 AC cycles and AC losses in each domain are calculated by spatially integrating $\boldsymbol{J} \cdot \boldsymbol{E}$ for the last half cycle.

$$Q = 2 \int_{1/f}^{3/(2f)} \int_{\Omega} \boldsymbol{J} \cdot \boldsymbol{E} d\Omega \, dt, \quad (15)$$

where $\Omega$ represents the superconducting domain and $f$ denotes the frequency of the applied AC source.

## 3. Results

Our analysis focuses on the geometric cases in which AC losses are expected to be the highest and the lowest: (1) the fully-aligned arrangement ($\beta = 0°$), wherein the enhanced perpendicular magnetic field should generate the highest AC losses and (2) the fully-misaligned arrangement ($\beta = 22.5°$), wherein the reduced perpendicular magnetic field should generate the lowest AC losses. When $\beta$ is varied from $0°$ to $22.5°$, AC losses are expected to vary between the values of the two extreme cases; Section 3.5 presents that data. All simulations assume good current distribution (i.e., the inner and outer layers carry the same transport current), but for those in Section 3.3 where the impact of current distribution on AC losses is presented. Also, most of the results were calculated at standard frequency $f = 50$ Hz, except for the simulations in Section 3.6 where the impact of frequency on AC losses is presented.

The parameters used in the paper are described in TABLE II.

TABLE II. Description of parameters used in the paper.

| Parameter | Description |
|---|---|
| $t_{HTS}$ | Thickness of HTS layer |
| gap | Size of inter-tape gap |
| $Q_{edge}$ or $Q_{1D}$ | AC loss generated near the edges by perpendicular fields |
| $Q_{surface}$ | AC loss generated by parallel magnetic fields |
| $Q_{total}$ or $Q_{2D}$ | Total AC losses calculated from 2D models |
| $I_{inner}$ | Current in the inner layer |
| $I_{outer}$ | Current in the outer layer |
| $I$ | Total current of the cable |

### 3.1. Effect of the $t_{HTS}$ on AC losses

Fig. 4 provides a comparison of edge loss $Q_{edge} = Q_{1D}$ (calculated using 1D models) and total losses, $Q_{total} = Q_{2D}$

(obtained from 2D models) in Cable 1, with $t_{HTS}$ varying from 1.5 μm to 10 μm. We simulate both the fully-aligned and fully-misaligned cases. We set a gap of 1 mm. As expected, for the same thickness of HTS layers, the AC losses generated in the fully-aligned arrangement are significantly higher (more than double) than those generated in the misaligned case, indicating that the contribution of the edge losses is more significant in the fully-aligned arrangement. The contribution of edge losses is expected to increase with increasing gap and the effect of the gap on AC losses will be further discussed in Section 3.2.

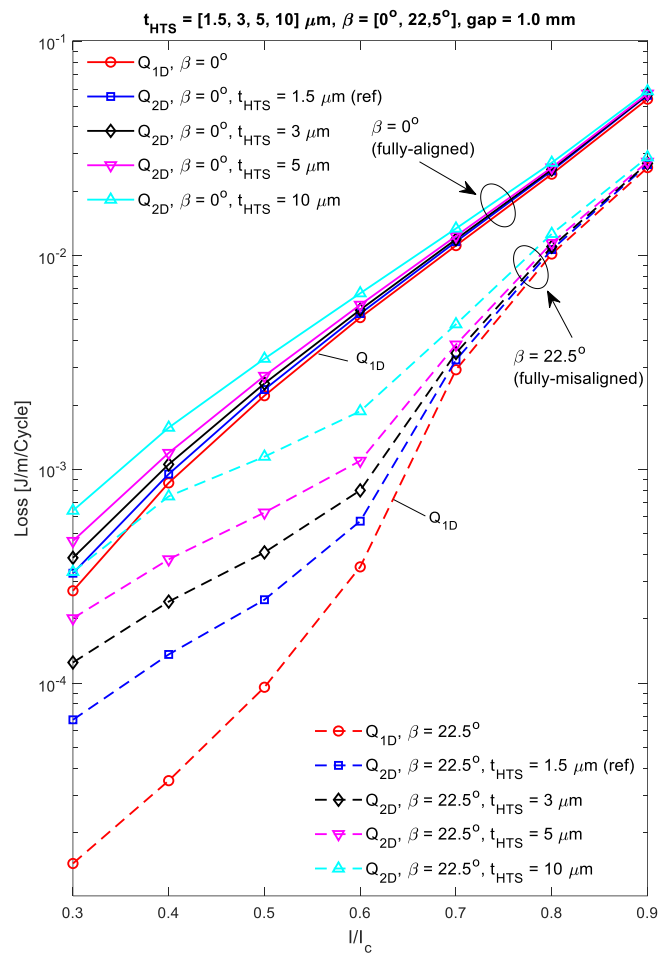

Fig. 4. The edge losses (calculated using 1D model) and total losses (obtained from 2D model) in Cable 1 with $t_{HTS}$ varying from 1.5 μm to 10 μm. Losses are plotted against the normalized current $I/I_c$ for both cases, $\beta = 0°$ (fully-aligned) and $\beta = 22.5°$ (fully-misaligned).

Because the 1D model only captures the edge AC losses, the difference between AC losses calculated by our 2D models and 1D models is the surface loss, $Q_{surface} = Q_{2D} - Q_{1D}$. The contribution of $Q_{surface}$ increases with the thickness of the HTS layers, especially for the case of fully-misaligned arrangement, $\beta = 22.5°$. To further analyze the effect of the $t_{HTS}$ on the contribution of the $Q_{surface}$ and $Q_{edge}$, AC losses plotted in Fig. 4 are normalized to $Q_{2D-1.5\mu m}$, the total loss calculated for a cable with $t_{HTS} = 1.5$ µm, close to the actual thickness of HTS layers in standard commercial tapes. The normalized losses, $Q/Q_{2D-1.5\mu m}$, are plotted in Figs. 5(a) and 5(b) for the fully-aligned case and fully-misaligned case, respectively. For the fully-aligned case (Fig. 5(a)), the 1D model underestimates the loss by around 20% when $I/I_c$ is equal to 0.3. However, when $I/I_c$ exceeds 0.6, the 1D model only underestimates the loss by 5%. Artificially increasing the $t_{HTS}$ results in a considerable overestimation of AC loss. For instance, if the $t_{HTS}$ increases from 1.5 µm to 10 µm, AC loss increases by nearly 200% at $I/I_c = 0.3$, by ~40% at $I/I_c = 0.5$ and by ~5% at $I/I_c = 0.9$.

µm). Losses are plotted against the normalized current $I/I_c$ for two cases: (a) $\beta = 0°$ and (b) $\beta = 22.5°$.

For the fully-misaligned case (Fig. 5(b)), the effect of $t_{HTS}$ on the cable AC loss is more significant. The 1D model underestimates the loss by around 80% when $I/I_c$ is equal to 0.3. However, when the $I/I_c$ exceeds 0.7, the difference in loss estimation reduces to less than 10% (Fig. 5b). Artificial increases of $t_{HTS}$ significantly overestimate AC loss in the cable. For instance, increasing $t_{HTS}$ from 1.5 µm to 10 µm results in an increase in cable AC losses by 500 % at $I/I_c = 0.3$, by ~330% at $I/I_c = 0.6$ and by ~10% at $I/I_c = 0.9$.

In short, when the applied transport current $I/I_c > 0.7$, the faster 1D model, with an underestimation of 10% or less, is a reasonable approximation. Moreover, artificially increasing the $t_{HTS}$ to reduce computational workload in the 2D simulations should be avoided, as this approach introduces a significant overestimation of AC losses, particularly as the gap gets smaller (Section 3.2 below).

### 3.2. Effect of the gap on cable AC losses

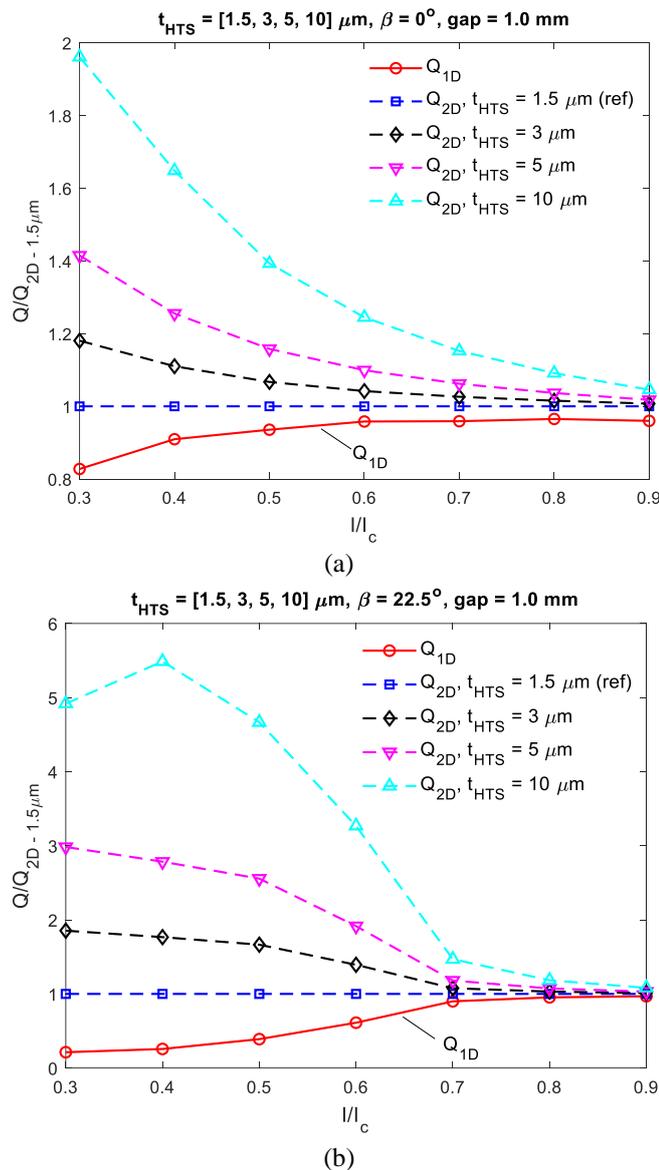

Fig. 5. Normalized AC losses $Q/Q_{2D-1.5\mu m}$, where Q are AC losses plotted in Fig. 4 and $Q_{2D-1.5\mu m}$ is the total loss calculated by 2D model for a cable of standard YBCO tapes ($t_{HTS} = 1.5$

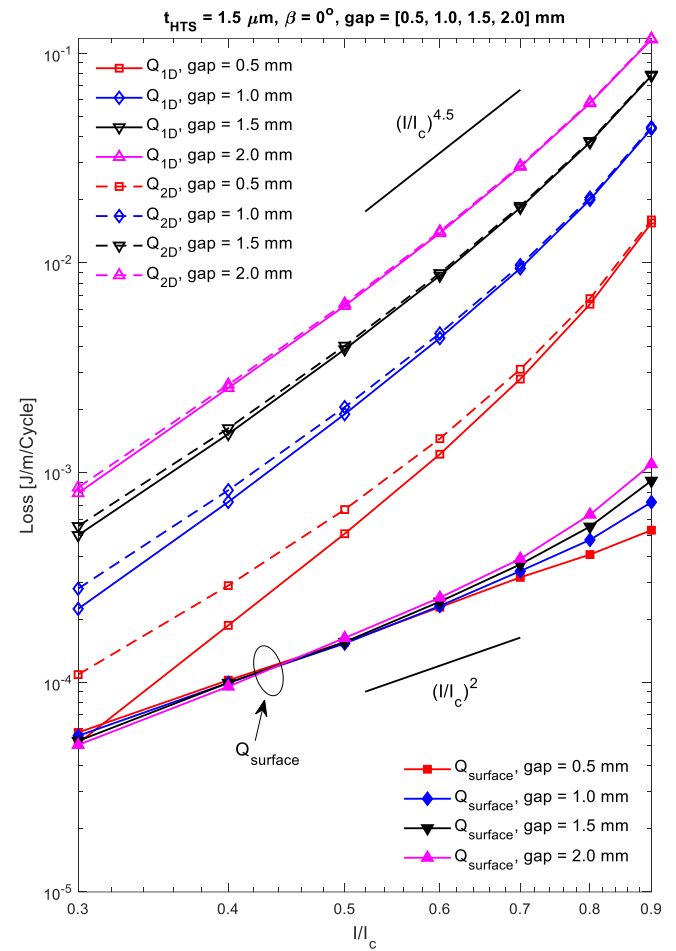

Fig. 6. AC losses obtained from 1D and 2D models are plotted against the normalized current $I/I_c$ for the case of $\beta = 0°$.

Fig. 6 illustrates the effect of various values of gap ranging from 0.5 mm to 2 mm on cable AC losses for the case of fully-aligned arrangement. Results obtained from both 1D and 2D models, along with $Q_{edge} = Q_{2D} - Q_{1D}$ are all plotted in Fig. 6 to establish the contributions of $Q_{edge}$ and $Q_{surface}$ for various values of gap and transport current. In general, larger gaps lead to higher total losses, as expected. The total AC loss losses increase by about 4 to 7 times as the gap increase from 0.5 mm to 2 mm, depending on the value of the transport current. The contribution of $Q_{edge}$ is calculated by our 1D model is dominant and increases quickly with increasing transport current. $Q_{edge}$ is roughly proportional to $(I/I_c)^4$. For gaps larger than 1.5 mm and $I/I_c > 0.5$, edge loss contributes more than 90% of the total loss. The 1D model can be used to predict AC losses with acceptable error for these cases. For gap = 0.5mm, the contribution of $Q_{edge}$ increases from 50% at $I/I_c = 0.3$ to ~ 95% at $I/I_c = 0.9$ of the total losses.

slowly compared to $Q_{edge}$ because $Q_{surface}$ is roughly proportional to $(I/I_c)^2$. Based on the results plotted in the figure, the contribution of $Q_{surface}$ is only larger than ~10% of the total loss when the gap is small (<0.5 mm) and at low current ($I/I_c < 0.6$). The 2D model is therefore preferable.

Fig. 7 depicts the effect of various gap values ranging from 0.5 mm to 2 mm on cable AC losses for the fully-misaligned configuration. Results obtained from both 1D and 2D models along with $Q_{surface} = Q_{2D} - Q_{1D}$ are also plotted. Again, the $Q_{surface}$ is nearly independent of the gap for quite a wide range of transport currents, $I/I_c < 0.75$. The slopes of both $Q_{edge}$ and $Q_{surface}$ curves change with the transport current. However, in general, $Q_{edge}$ increases with transport current at a much faster rate than $Q_{surface}$.

For low currents, the AC losses calculated by the 2D model are significantly higher than those calculated by the 1D model, indicating that $Q_{surface}$ is dominant. Thus, cable AC losses are independent of the gap when $I/I_c < 0.5$. At higher currents, the contribution from $Q_{edge}$ becomes more significant when the current increases--the 1D results approach the results obtained from 2D model. Hence, the effect of the gap size on the cable AC losses becomes more significant as current increases. At $I/I_c = 0.9$, the cable total AC losses increase 600% when the gap increase from 0.5 mm to 2 mm. In conclusion, the effect of gap size on cable AC losses for the fully-misaligned case strongly depends on the transport current, which determines the relative contribution of the surface and edge losses.

### 3.3. Effect of the current distribution on AC losses.

Cables are usually designed to have perfectly balanced current distribution between the layers. However, this condition can be difficult to achieve, and cables may operate with unbalanced current distribution [1,8]. Establishing the relationship between layer current distribution and loss is therefore useful. Fig. 8 illustrates the analysis of AC losses obtained from the 2D model for different transport current distribution ratios, $I_{inner}/I$. Here $I_{inner}$ is the current in the inner layer and I is the total current in the cable. We compute losses for three values of gap, gap = 0.5, 1 and 1.5 mm, and three values of the transport current, $I = 0.5I_c$, $0.7I_c$, and $0.9I_c$. Results for both the fully-aligned and fully-misaligned cases are depicted in Fig. 8(a) and 8(b), respectively. In general, the lowest AC losses are identified at or near $I_{inner}/I = 0.5$ (well-balanced current sharing) for all the data in both figures. The effect of the off-balanced current sharing on total AC losses is stronger at the higher cable transport currents and smaller gaps, no matter whether the inner layer carries either more or less than the outer layers. Similar behaviors have been observed experimentally in [33-34].

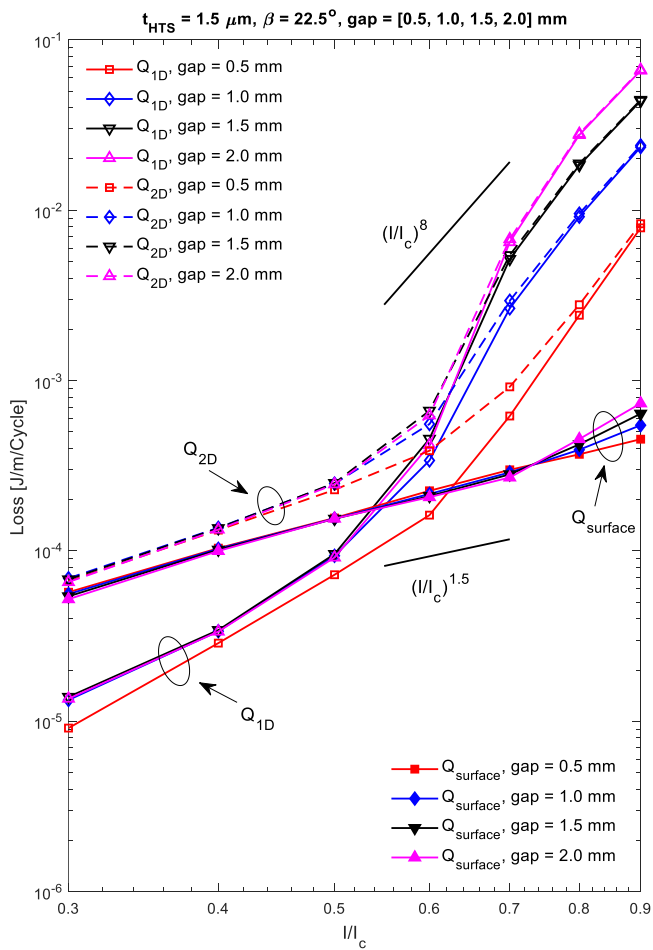

Fig. 7. AC losses obtained from 1D and 2D models are plotted against the normalized current $I/I_c$ for the case of $\beta = 22.5°$.

The surface loss, $Q_{surface}$, is nearly independent of the size of gap at $I/I_c < 0.6$. $Q_{surface}$ slightly increases for larger gaps at high currents, $I/I_c < 0.6$. $Q_{surface}$ increases with current at a

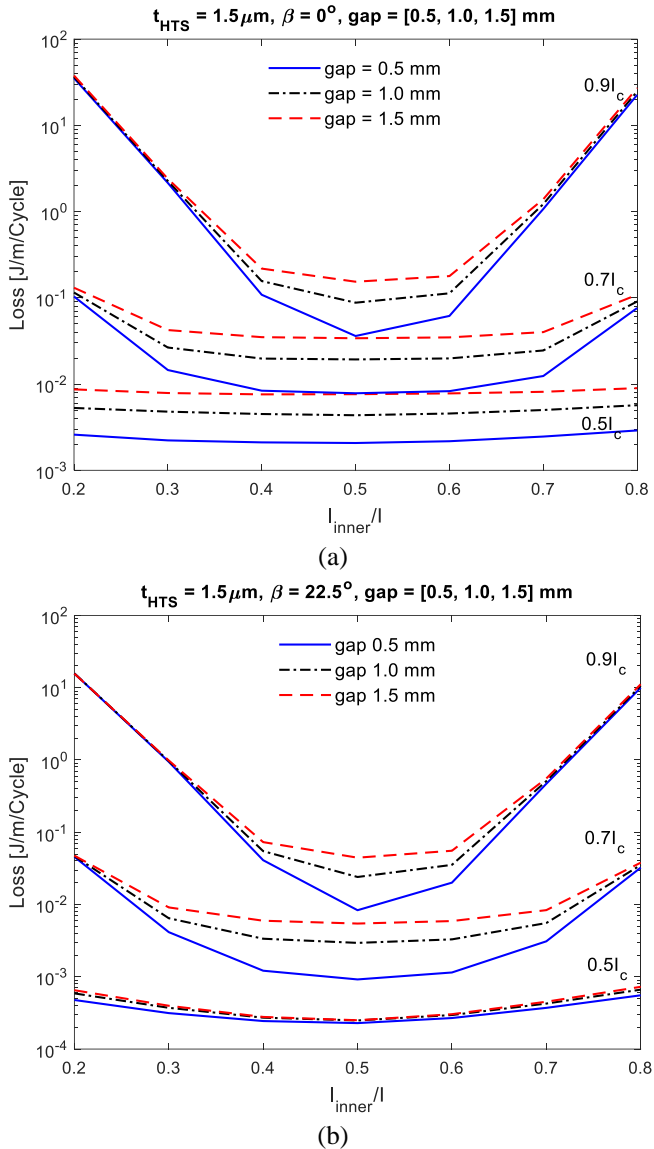

(a)

(b)

Fig. 8. AC losses in Cable 1 as a function of the transport current distribution ratio I$_{inner}$/I for both cases, (a) β = 0° and (b) β = 22.5°.

### 3.4. Individual contribution of AC loss in the inner and outer layers

Despite the inner and outer layers carrying balanced current, they are expected to experience distinct magnetic fields and generate varying AC losses. Figure 9 illustrates the AC loss curves for both tape arrangements, obtained from either the 1D model (Fig. 9a) or the 2D model (Fig. 9b). The results from the 1D model depicted in Fig. 9a show that the inner and outer layers generate nearly same edge loss for either β = 0° or β = 22.5°. Conversely, the results depicted in Fig. 9b demonstrate that the outer layer generates slightly higher AC losses in both the fully-aligned and fully-misaligned cases. Nonetheless, the discrepancy in AC loss is more notable at lower I/I$_c$s, due to the dominance of surface loss in the low current region, as discussed in Section 3.2. This outcome is to be expected, as the outer layer experiences higher parallel fields, yielding higher surface losses. To confirm the results shown in Fig. 9 and visualize the electrodynamics of the HTS tapes, the current profiles along the half-width of HTS tapes in both inner and outer layers at several points during an AC cycle are depicted in Fig. 10. The current profiles are plotted at 6 points labeled 1 to 6, as seen in the inserts of those figures. The evolutions of the current distribution are nearly the same for the inner and outer layers. At each time step, the current density in the outer layer seems to be slightly higher, generating higher AC losses as seen in Fig. 9. The current density reaches or slightly surpasses the Jc(B) in small regions near the gaps where experience the application of the perpendicular magnetic field.

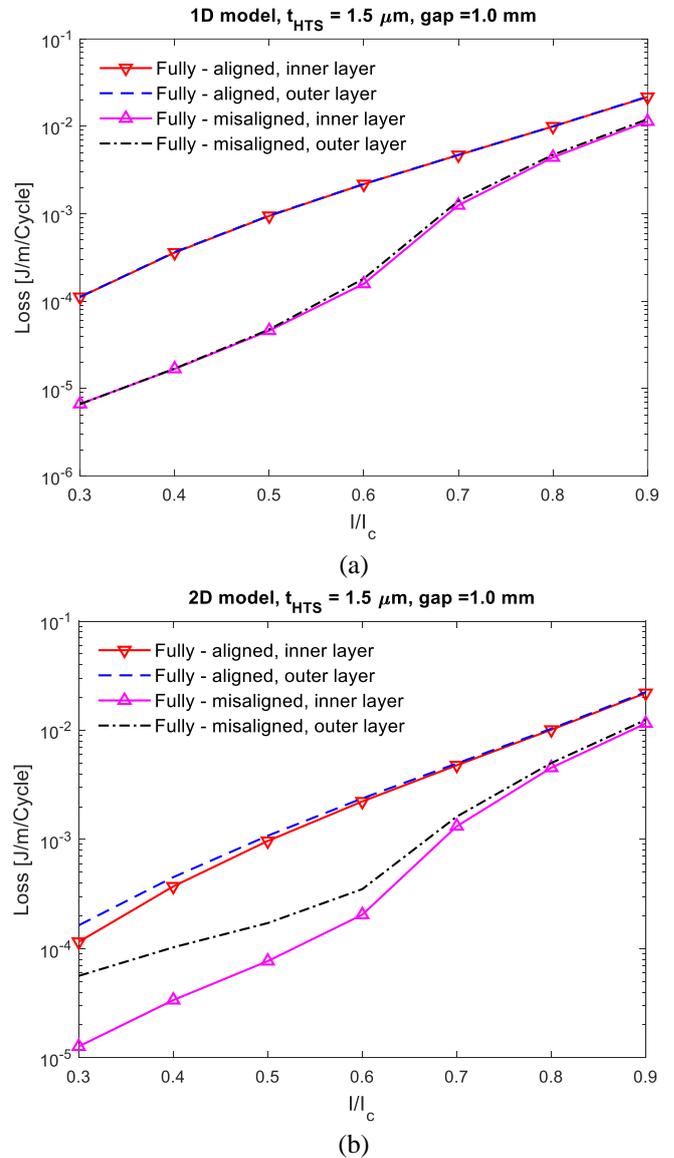

(a)

(b)

Fig. 9. AC losses generated in the inner and outer layers as functions of the normalized current I/I$_c$ for calculated by (a) 1D model and (b) 2D model.

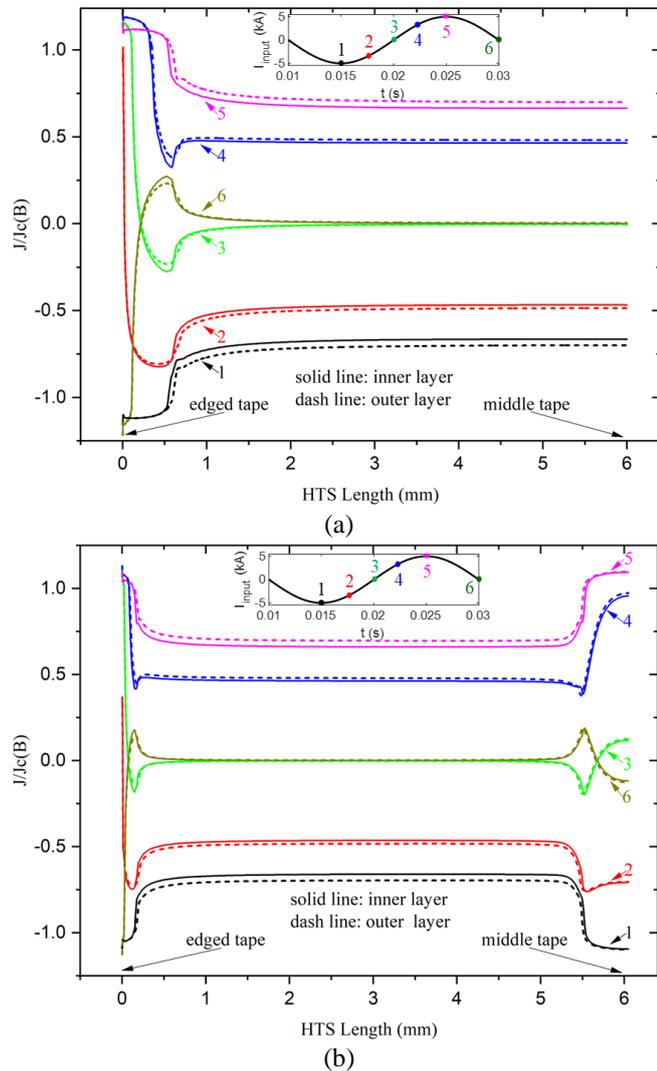

Fig. 10. Distribution of the normalized current density $J/J_c(B)$ along the half-width of HTS tapes in both inner and outer layers at several points during an AC cycle for (a) $\beta = 0°$ and (b) $\beta = 22.5°$.

### 3.5. Effect of the HTS tape arrangement on AC losses.

To achieve balanced cables, the inner and outer layers are wound in opposite directions. Therefore, angle $\beta$ varies from 0° to 22.5° as discussed prior. Fig. 11 plots the dependence of the total AC losses on $\beta$ for several values of transport current and gap for $t_{HTS} = 1.5$ µm. Generally, the AC losses are the highest near $\beta = 0°$ and the lowest when $\beta = 22.5°$, as expected. For gap = 0.5 mm, AC losses quickly decrease when $\beta$ increases, and the AC losses are minimal for $\beta > 5°$; perpendicular fields are only present near gaps, and magnetic flux lines quickly align to the tapes' surface away from those gaps. Hence magnetic fields realized on the HTS tapes are nearly identical for $\beta$ between 5° to 22.5°.

For larger gaps, effect of $\beta$ on the total AC losses is similar. However, for those larger gaps, maximal AC losses are observed when $\beta$ is a few degrees shifted from 0°, suggesting a higher magnetic field in this configuration.

The results in Fig. 11 suggested a possible improvement in predicting the total AC losses in CORC cables. While the calculation of AC losses at $\beta = 0°$ and $\beta = 22.5°$ can basically provide the highest and lowest values, the average AC losses calculated from the curves in Fig. 11 for the entire range of $\beta$ would provide a better estimation of AC losses in cables.

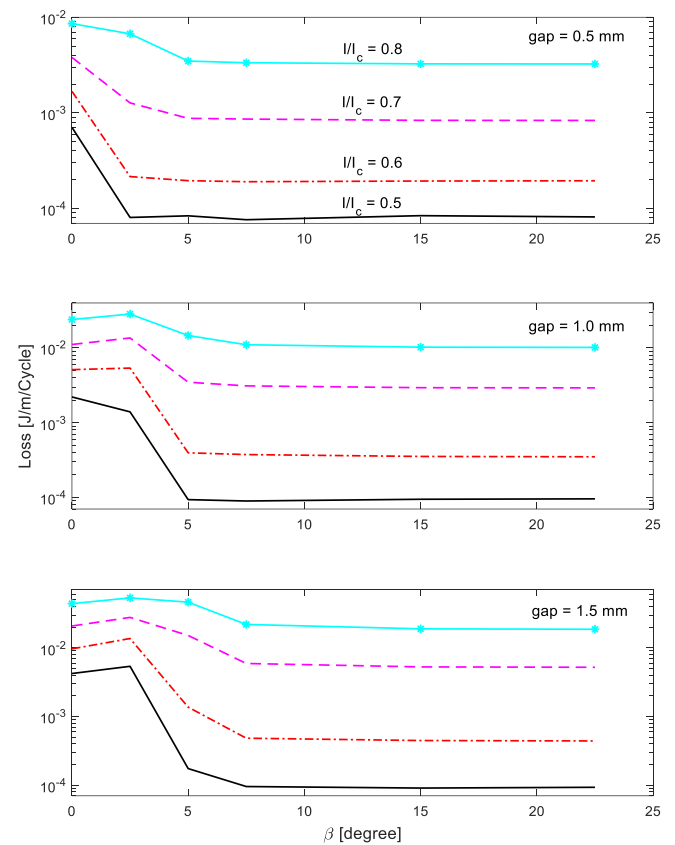

Fig. 11. Total AC loss in cable 1 as functions of $\beta$ for different gaps (gap = 0.5 mm, 1 mm, and 1.5 mm) and at different values of the transport current ($I = 0.5I_c$, $0.6I_c$, $0.7I_c$ and $0.8I_c$).

### 3.6. Effect of frequency on AC loss in cables

The AC losses in HTS originates from (1) flux-creep loss, which is hysteretic in nature and dominates at for low transport currents, and (2) flux-flow loss, which is resistive in nature and dominates for high transport currents. Induced currents generated in normal metal layers (substrates and stabilizers) also generate the eddy current losses. Because these three losses depend differently on frequency, we simulate three frequencies of 50, 250, and 500 Hz.

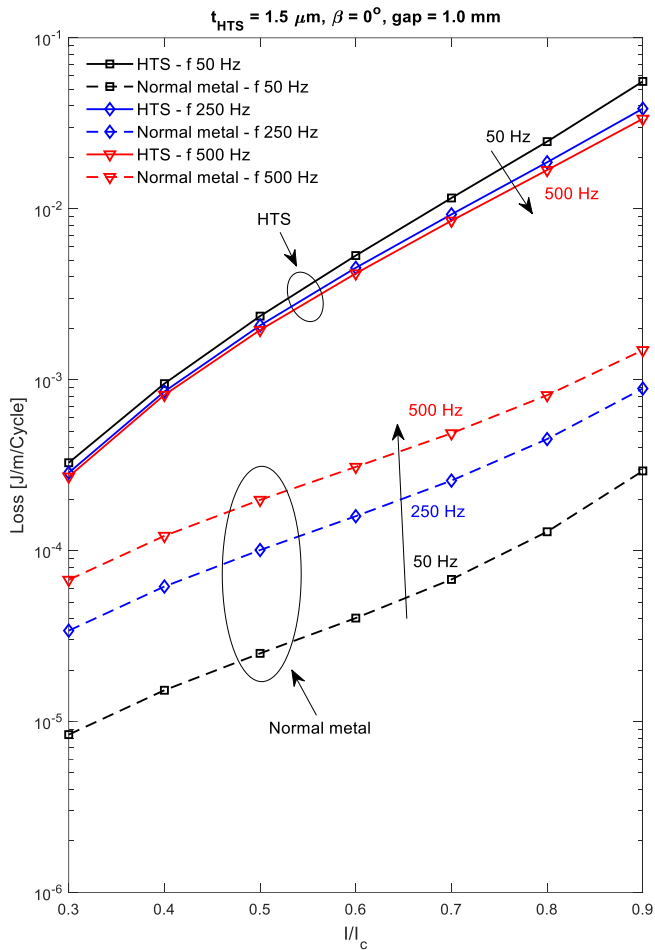

Fig. 12. AC losses in the HTS layers and normal metal layers at 50 Hz, 250 Hz and 500 Hz for the fully-aligned case

Figure 12 plots the total AC losses per cycle in the HTS and non-superconducting layers at 50 Hz, 250 Hz, and 500 Hz for the fully-aligned case. Generally, the losses in normal metal layers are significantly lower than those generated in HTS layers. The eddy current loss in normal metal per an AC cycle seems to be proportional to the frequency as expected. The AC losses in the HTS layer slightly decrease with increasing frequency and that decrease becomes more considerable at higher currents. This behavior suggests that there may be a contribution of the flux-flow loss (which is proportional to $1/f$,) and that contribution is greater with increasing transport current as expected.

Figure 13 plots AC losses in the HTS and normal metal layers at 50 Hz, 250 Hz and 500 Hz for the case of $\beta = 22.5°$. Again, the eddy current loss per cycle of the normal metal layers also increases proportionally to frequency. Though the HTS losses are lower for this case ($\beta = 22.5°$) in comparison to the case of $\beta = 0°$, they are still higher than the eddy current losses when $I/I_c > 0.6$. The HTS losses are independent of frequency for $I/I_c < 0.5$, implying that flux creep loss is dominant. The situation is changes drastically when the transport current increases to higher values. The HTS AC loss decreases with increasing frequency for $I/I_c > 0.7$.

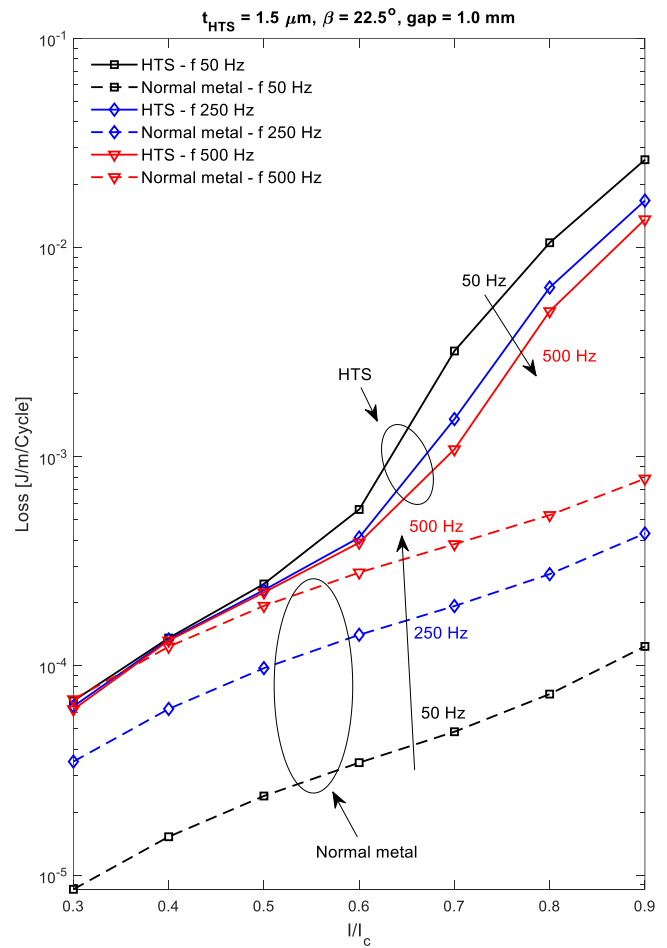

Fig. 13. AC losses in the HTS layers and normal metal layers at 50 Hz, 250 Hz and 500 Hz for the fully-misaligned case,

To make explicit the effect of frequency on the AC losses in the HTS, the perpendicular magnetic fields applied along the half-width of HTS tapes are calculated at the positive peak current of $I = 0.7I_c$ and are plotted in Fig. 14 (for the case $\beta = 22.5°$) and Fig. 15 (for the case $\beta = 0°$). The half-width of the HTS tape of the inner layer is denoted as arc-length AB and the half-width of the HTS tape of the outer layer is denoted as arc-length CD (see Figs. 14 and 15). As expected, the perpendicular fields are only non-zero at small regions very close to the gaps. The rest of the tape width experiences no perpendicular fields. In both Figs. 14 and 15, the perpendicular magnetic fields are slightly higher when the frequency is lower, causing higher AC loss at lower frequencies, as seen in Figs 12 and 13. There is no elucidative explanation for this observation, but the shielding effect of the HTS layers and metal layers might contribute.

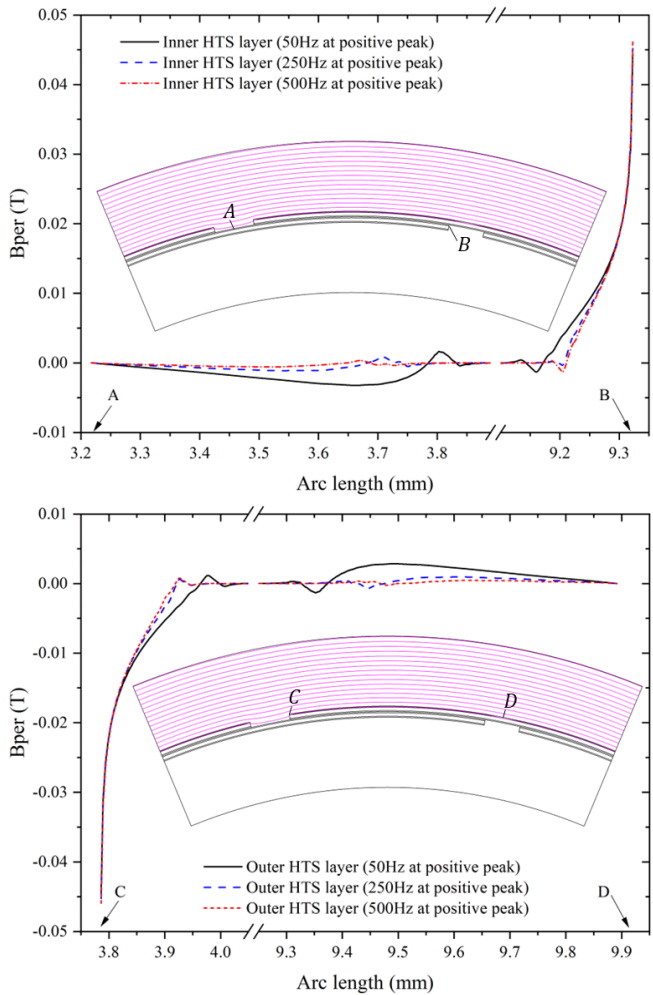

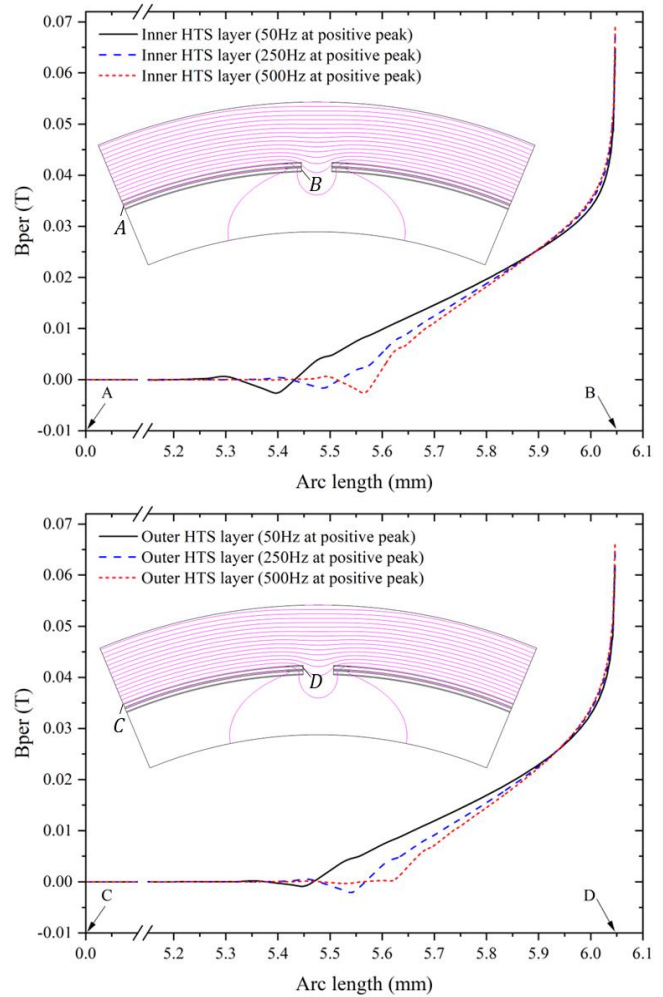

edge loss is dominant. For this case, AC loss in Cable 2 is about 60% higher for nearly the entire current range.

Fig. 14. Perpendicular magnetic flux density distribution along the half-width of HTS tapes in the inner and outer HTS layers of a fully-misaligned cable. The cable carries a transport current $I/I_c = 0.7$ at various current frequencies: 50, 250, and 500 Hz.

### 3.7. Effect of number of gaps on AC loss in cables

Coated conductors are now commercially available in several widths. When narrower tapes are used to make a cable, the number of tapes (and therefore the number of gaps) must be increased to achieve the same cable $I_c$. This section compares AC losses in two two-layer cables of the same diameter, $I_c$, and gap size, the first fabricated from 16 tapes of 12.5 mm width and the other fabricated from 32 tapes of 5.75 mm width. Detailed specifications can be found in TABLE I. The cables have the same winding diameter, so their circumferential fields should have the same amplitude if the cables carry the same transport current. Figure 16 plots transport AC losses in Cables 1 and 2 for both cases, $\beta = 0°$ and $\beta = 22.5°$. The AC losses in Cable 2 are higher than those generated in Cable 1, especially in the case of $\beta = 0°$ when the

Fig. 15. Perpendicular magnetic flux density distribution along the half-width of HTS tapes in the inner and outer HTS layers of a fully-aligned cable. The cable carries a transport current $I/I_c = 0.7$ at various current frequencies: 50, 250, and 500 Hz.

## 4. Concluding Remarks and Discussions

We have provided insights into the behavior of the AC losses of CORC HTS cables by utilizing both 1D and 2D models. Our results can be summarized as follows:

1) The AC losses are highest when $\beta = 0°$ (the gaps of the inner and outer layers are perfectly aligned). The AC losses decrease and flatten when gaps are not aligned ($5° < \beta < 22.5°$). Because the latter represents a greater angular range, the losses calculated for misaligned gaps represent better the AC losses in the cables.

2) AC losses in HTS cable are the result of two components: the surface loss generated by parallel

circumferential field and edge loss generated by the perpendicular fields generated locally near the gaps. For configurations with misaligned gaps, the surface loss is dominant for the low current region $I/I_c < 0.5$. However, the edge loss increases quickly with increasing current and becomes a dominant contribution at $I/I_c > 0.6$.

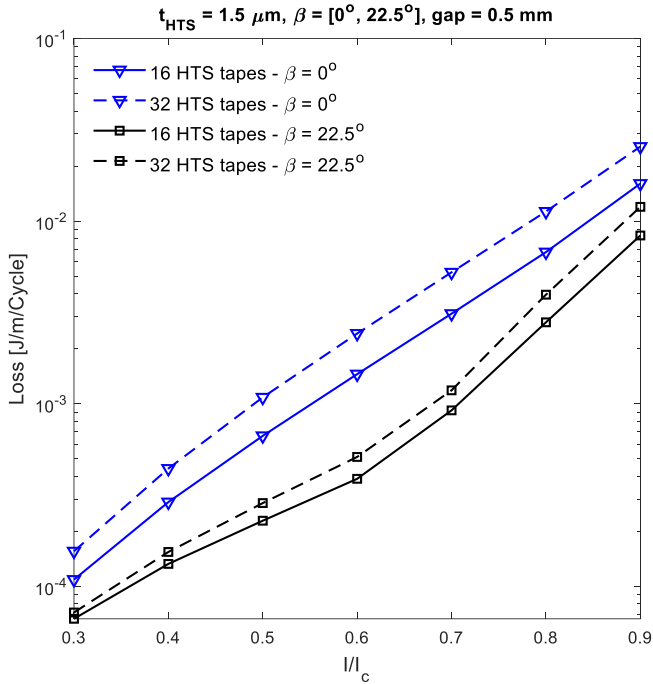

Fig. 16. Transport AC losses in Cables 1 and 2 for $\beta = 0°$ and $\beta = 22.5°$. Both cables have the same diameter, $I_c$ and gap size, but one fabricated by 16 tapes of 12.5 mm width and the other fabricated by 32 tapes of 5.75 mm width. Detailed specifications can be found in TABLE I.

3) Because the edge loss, which strongly depends on the gap size, becomes the dominant contribution for high transport currents, reducing the number of gaps by using wider HTS tape should reduce AC losses significantly. However, it is challenging to conform a wider HTS tape to a small former. Flexible, wide HTS tapes with thin substrates would be beneficial for cables with lower AC losses.

4) Gap size should be minimized to reduce AC losses. Besides the low AC loss target, cables are also designed to have a balanced current-sharing operation with the lowest cable inductance (or axial magnetic field). These objectives - the smallest gaps, balanced current sharing, and low cable inductance - are difficult to achieve simultaneously, but the availability of HTS tapes at a larger variety of widths should ease the process.

5) In the operating current region ($I/I_c > 0.5$) and frequencies below 500 Hz, AC losses generated in the HTS layer are still dominant. HTS losses (per cycle) decrease as the frequency increases; the perpendicular magnetic fields near gaps decrease when the frequency increases. The shielding effects of the HTS layers and metal layers might produce this effect.

## Acknowledgements

This work was supported in part by the ARPA-E of DoE under Contract DE-FOA-0002459 and in part by NSF under Grants DMR-1644779 and DMR-2128556.

## References

[1] Weiss JD *et al* 2017 *Supercond. Sci. Technol.* **30** 014002
[2] Yuan W *et al* 2018 *IEEE Trans. Appl. Supercond.* **28** 5401505
[3] Lee SJ *et al* 2018 *IEEE Trans. Appl. Supercond.* **28** 5401205
[4] Lee C *et al* 2020 *Supercond. Sci. Technol.* **33** 044006
[5] Nguyen LN *et al* 2023 *IEEE Trans. Appl. Supercond.* **33** 5600208
[6] Sumption MD *et al* 2022 *Supercond. Sci. Technol.* **35** 025006
[7] Terzioğlu R *et al* 2017 *Supercond. Sci. Technol.* **30** 085012
[8] Ogawa J *et al* 2020 *IEEE Trans. Appl. Supercond.* **30** 5900205
[9] Yamaguchi T *et al* 2014 *SEI Tech. Rev.* **78** 79-85. *global-sei.com/technology/tr/bn78/pdf/78-16.pdf*
[10] Ashworth SP and Nguyen DN 2010 *Supercond. Sci. Technol.* **23** 095009
[11] Han J *et al* 2022 *IEEE Trans. Appl. Supercond.* **32** 5900705
[12] Nguyen DA *et al* 2010 *IEEE Trans. Appl. Supercond.* **21** 996-1000.
[13] Yagotintsev K *et al* 2020 *Supercond. Sci. Technol.* **33** 085009
[14] Malozemoff AP *et al* 2009 *IEEE Trans. Appl. Supercond.* **19** 3115-3118.
[15] Grilli F *et al* 2013 *IEEE Trans. Appl. Supercond.* **24** 8200433.
[16] Siahrang M *et al* 2012 *Supercond. Sci. Technol.* **25** 014001
[17] Grilli F *et al* 2010 *Supercond. Sci. Technol.* **23** 034017
[18] Zermeno VM *et al* 2013 *Supercond. Sci. Technol.* **26** 052001
[19] Shen B *et al* 2020 *Supercond. Sci. Technol.* **33** 033002
[20] Shen B 2021 *IEEE Trans. Appl. Supercond.* **31** 4803505
[21] Fu S *et al* 2018 *IEEE Trans. Appl. Supercond.* **28** 4802005
[22] Siahrang M *et al* 2010 *IEEE Trans. Appl. Supercond.* **20** 2381-2389.
[23] Huber F *et al* 2022 *Supercond. Sci. Technol.* **35** 043003
[24] Wang Y *et al* 2020 *Physica C* **579** 1353770. *doi.org/10.1016/j.physc.2020.1353770*
[25] Yang J *et al* 2021 *IEEE Trans. Appl. Supercond.* **31** 5901504
[26] Wang S *et al* 2022 *Supercond. Sci. Technol.* **35** 065013
[27] Li Q *et al* 2023 *IEEE Trans. Appl. Supercond.* **33** 4800207
[28] Shen B *et al* 2021 *IEEE Trans. Appl. Supercond.* **31** 4803405
[29] Sato S and Amemiya N 2006 *IEEE Trans. Appl. Supercond.* **16** 127-130
[30] Li Q *et al* 2013 *Physica C* **484** 217-222. *doi.org/10.1016/j.physc.2012.02.033*
[31] Yang J *et al* 2021 *J. Supercond. Nov. Mag.* **35** 57-63. doi.org/10.1007/s10948-021-06031-5
[32] Amemiya N *et al* 2007 *IEEE Trans. Appl. Supercond.* **17** 1712-1717

[33] Ogawa J *et al* 2018 *IEEE Trans. Appl. Supercond.* **28** 5900104
[34] Ogawa J *et al* 2011 *IEEE Trans. Appl. Supercond.* **22** 4704804
[35] www.advancedconductor.com
[36] www.comsol.com
[37] Grilli F *et al* 2014 *IEEE Trans. Appl. Supercond.* **24** 8000508